\documentstyle[hx37_latex]{article}

\def\be{\begin{equation}}
\def\ee{\end{equation}}
\def\bea{\begin{eqnarray}}
\def\eea{\end{eqnarray}}

\begin{document}

\title{PHOTOMETRIC REDSHIFTS IN THE HUBBLE DEEP FIELD}

\author{ M.J. SAWICKI, H. LIN, H.K.C. YEE}

\address{Department of Astronomy, University of Toronto, Toronto, M5S 3H8, Canada}


\maketitle\abstracts{
Photometric redshifts for galaxies in the Hubble Deep Field are
measured.  
Luminosity functions show steepening of the faint-end 
slope and mild brightening of $M^*$ out to $z\approx 3$,
followed by a decline at higher $z$; 
an excess of faint, star-forming galaxies is seen at low $z$.  
Our results are consistent with the
formation of large galaxies at $z=2$--$3$, followed 
by that of dwarfs at $z<1$.}

\section{Colour Redshifts in the HDF}

Because of the extreme depth of the Hubble Deep Field (HDF),
spectroscopic redshifts are not practical for all but the brightest
objects.  
The redshift of a galaxy can, however, be estimated by
comparing the observed broadband colours against a set of reference
templates computed for a range of redshifts and spectral types.  
We computed templates by extending empirical spectral energy distributions
\cite{col80} into the UV, applying Lyman blanketing \cite{mad95} at high $z$, 
and convolving with HST filter transmission curves. 

Object finding and photometry was done using the PPP faint galaxy
photometry package \cite{yee91}.  There are 1003 objects with
F814W$_{AB} \leq 27$, of which 90\% are detected in all four HDF
bandpasses.  
Each object's observed colours were compared against the templates
(using a least-squares fitting technique) to obtain the most likely 
redshift and spectral type.
Photometric redshifts agree well with
spectroscopic ones, with a scatter of $\sigma_z = 0.13$ at $z<1.5$,
increasing to $\sigma_z = 0.32$ at $z>2$.  The catastrophic failure
rate is small (2/57 objects).

\section{Galaxy Population to $z=4$}

For a full discussion the reader is referred to our main paper
\cite{saw96}.  
The results are summarized in Fig.\ 1(a--c).  The bright 
end of the luminosity function (LF) brightens moderately between the
present epoch and $z\approx 2.5$.  
This brightening is accompanied by a steepening of the faint-end slope.  
Beyond $z\approx 3$ the LF fades to values similar to those seen locally. 
This fading could be a
signature of the onset of galaxy formation which is expected to occur
around that redshift~\cite{fuk96}.  
In this scenario, the star-forming
galaxies seen at $z>2$ in Fig.\ 1b will become present-day ellipticals
and spirals; those at $z\approx 1$ are star-forming dwarfs which are
also seen in the Hubble diagram, and evolve to become the faint 
(M$_{F450W_{AB}}>-15$) galaxies in the $z=0.2$--$0.5$ LF.

\newpage

\begin{figure}
\includegraphics{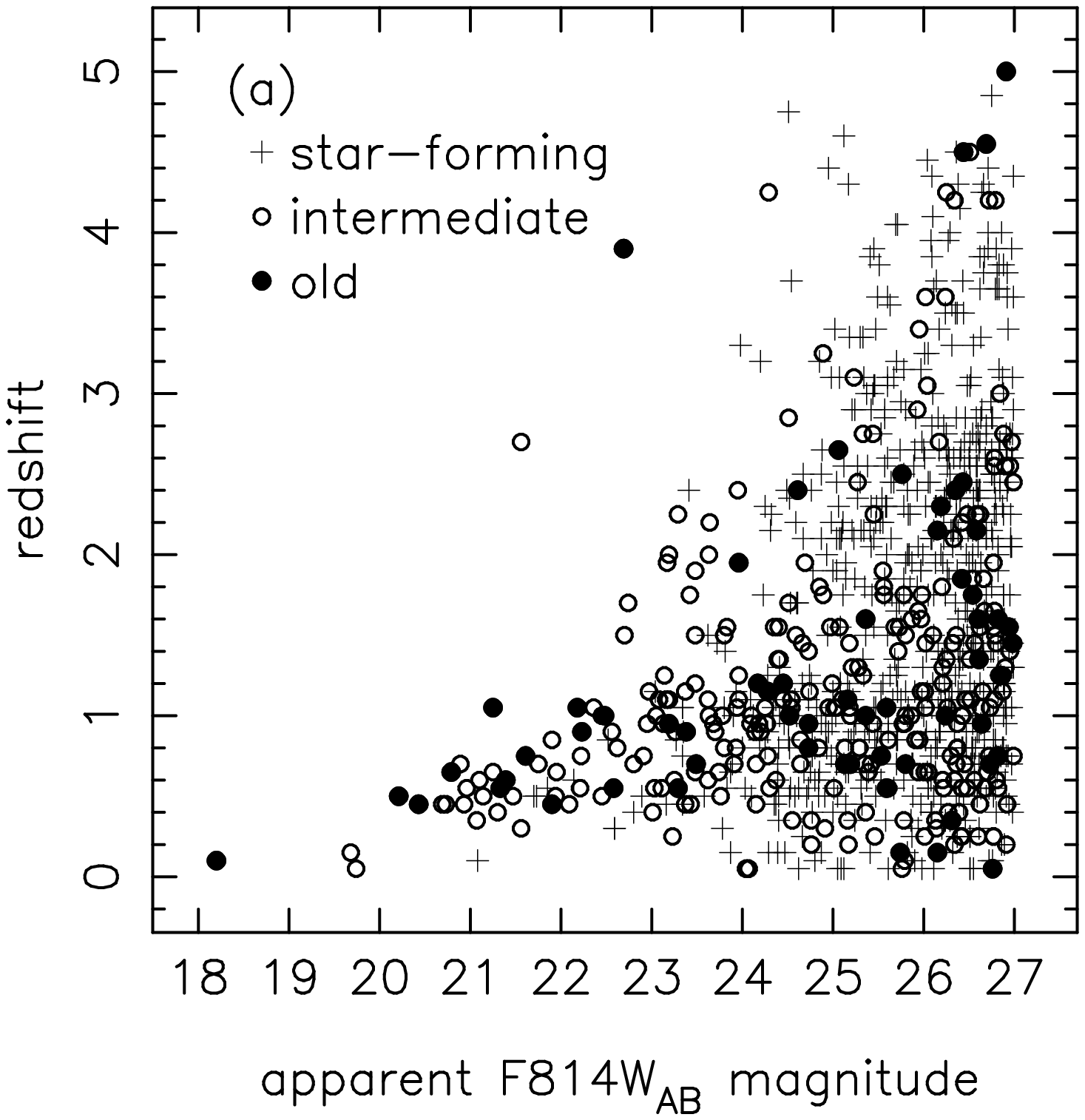}
\includegraphics{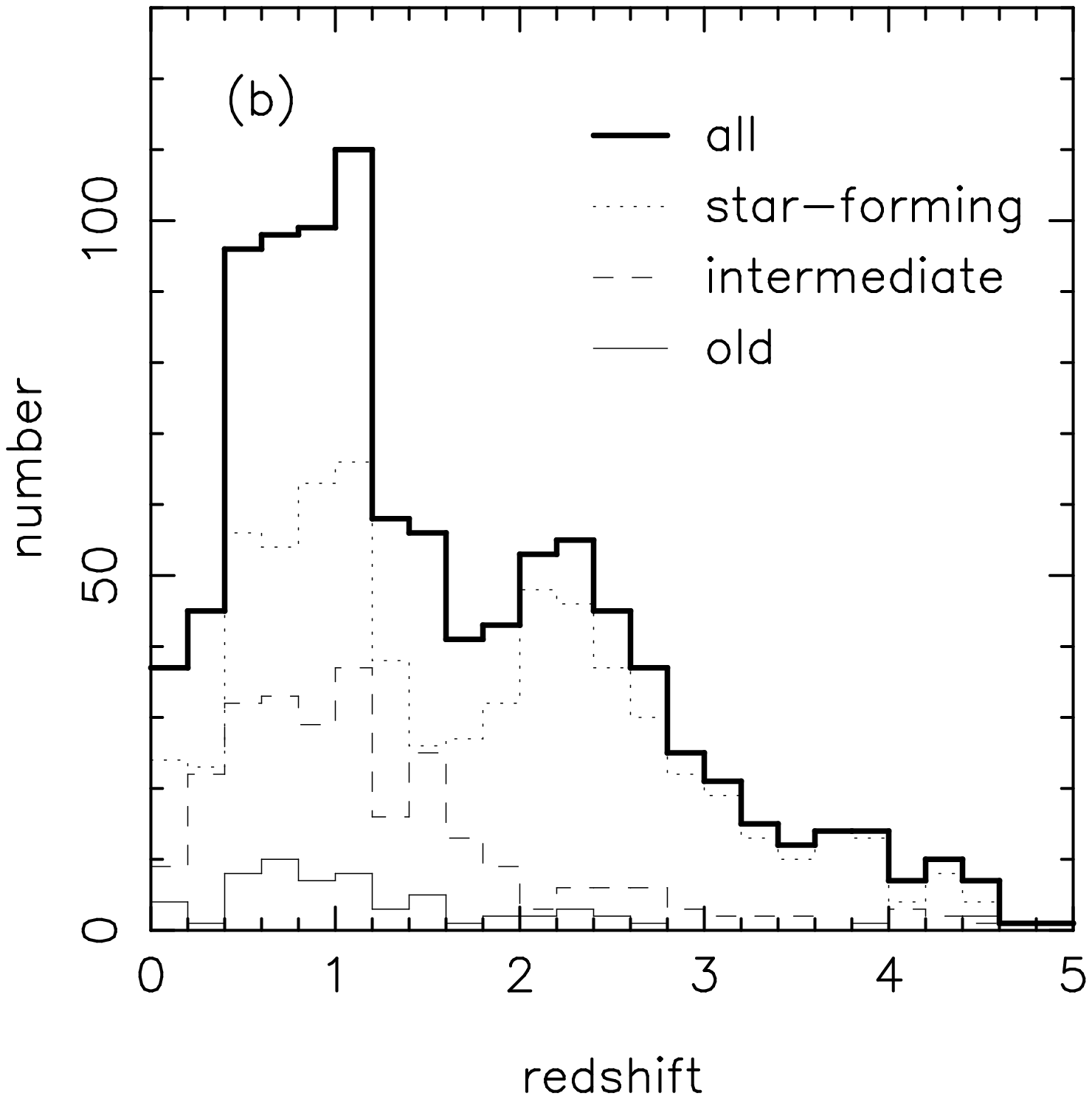}
\includegraphics{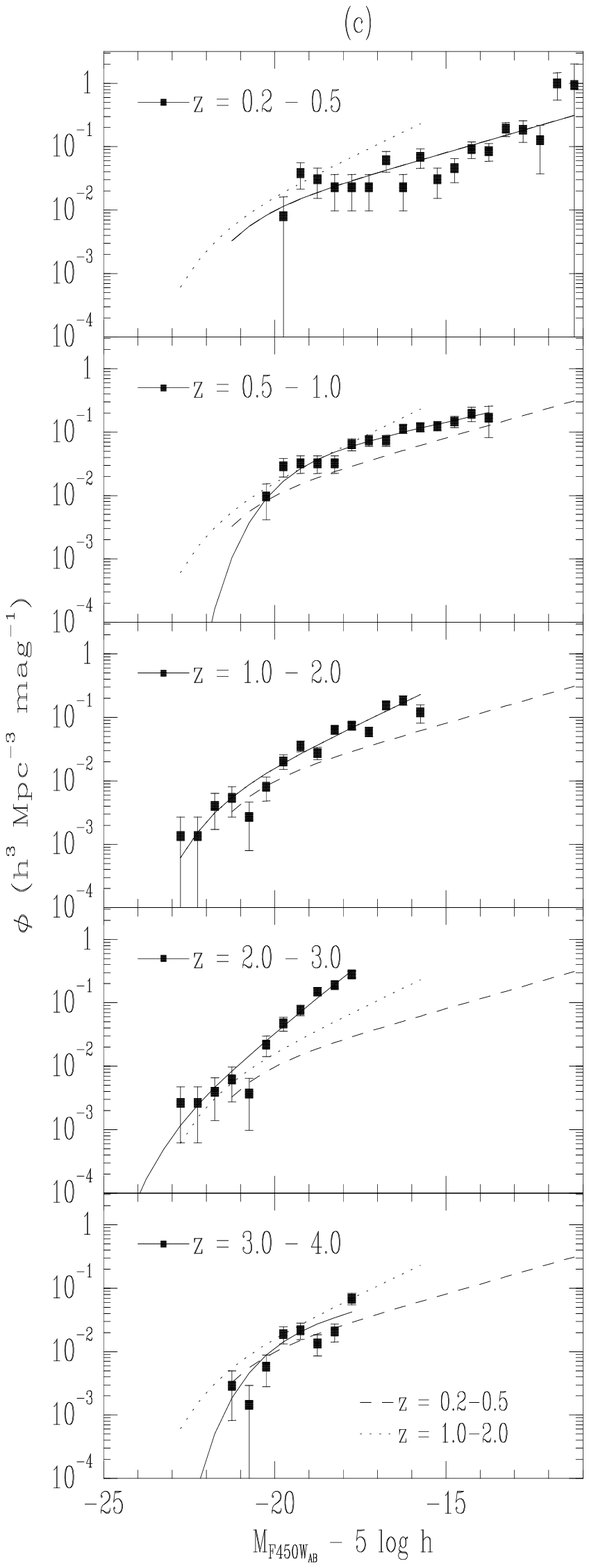}
\end{figure}

\begin{figure}
\vspace{10.24cm}
\caption{
(a) Hubble diagram. 
(b) Redshift distribution for objects with F814W$_{\rm AB}\leq 27$. 
(c) F450W$_{\rm AB} (\approx B_{\rm AB})$ luminosity functions;  
dashed and dotted lines are fiducials.}
\end{figure}

%
%

\end{document}